\documentclass[aps,prl,twocolumn,preprintnumbers,amsmath,amssymb,superscriptaddress]{revtex4-2}
\usepackage{graphicx}
\usepackage[caption=false]{subfig}
\usepackage{epsfig}
\usepackage{dcolumn}
\usepackage{bm}
\usepackage{color}
\usepackage{float}
\usepackage[colorlinks]{hyperref}
\usepackage{verbatim}
\usepackage{algorithm}
\usepackage{algorithmic}
\usepackage{tikz}
\usetikzlibrary{quantikz}

\hypersetup{citecolor = blue}

\def\be{\begin{equation}}       \def\ee{\end{equation}}
\def\bea{\begin{eqnarray}}      \def\eea{\end{eqnarray}}
\def\ba{\begin{array}}
\def\ea{\end{array}}
\def\bnum{\begin{enumerate} }
\def\enum{\end{enumerate}}

\def\=>{\Rightarrow}
\def\>{\rightarrow}

\def\eye2{Fathbb{I}}

\renewcommand{\>}{\rangle}

\renewcommand{\rm}[1]{\mathrm{#1}}

\usepackage{listings}
\usepackage{color}
\definecolor{lightgray}{gray}{1}

\begin{document}

\title{Training variational quantum algorithms with random gate activation}
\author{Shuo Liu}
\affiliation{Institute for Advanced Study, Tsinghua University, Beijing 100084, China}
\author{Shi-Xin Zhang}
\email{shixinzhang@tencent.com}
\affiliation{Tencent Quantum Laboratory, Tencent, Shenzhen, Guangdong 518057, China}
\author{Shao-Kai Jian}
\email{sjian@tulane.edu}
\affiliation{Department of Physics \& Engineering Physics, Tulane University, New Orleans, LA 70118, USA}
\author{Hong Yao}
\email{yaohong@tsinghua.edu.cn}
\affiliation{Institute for Advanced Study, Tsinghua University, Beijing 100084, China}

\begin{abstract}
Variational quantum algorithms (VQAs) hold great potentials for near-term applications and are promising to achieve quantum advantage on practical tasks. 
However, VQAs suffer from severe barren plateau problem as well as have a large probability of being trapped in local minima. 
In this Letter, we propose a novel training algorithm with random quantum gate activation for VQAs to efficiently address these two issues. 
This new algorithm processes effectively much fewer training parameters than the conventional plain optimization strategy, which efficiently mitigates barren plateaus with the same expressive capability.
Additionally, by randomly adding two-qubit gates to the circuit ansatz, the optimization trajectories can escape from local minima and reach the global minimum more frequently due to more sources of randomness. 
In real quantum experiments, the new training algorithm can also reduce the quantum computational resources required and be more quantum noise resilient. 
We apply our training algorithm to solve variational quantum simulation problems for ground states and present convincing results that showcase the advantages of our novel strategy where better performance is achieved by the combination of mitigating barren plateaus, escaping from local minima, and reducing the effect of quantum noises. 
We further propose that the entanglement phase transition could be one underlying reason why our RA training is so effective.
\end{abstract}

\date{\today}
\maketitle

{\bf Introduction:}
Recently, various quantum-classical hybrid variational algorithms, such as variational quantum eigensolver (VQE) \cite{peruzzo_VQE_2014,mcclean_VQE_2016}, quantum approximate optimization algorithm (QAOA) \cite{farhi_QAOA}, and quantum neural networks \cite{farhi_classification_2018_QNN}, have been proposed with the vision of establishing valuable quantum killer-apps in Noisy Intermediate-Scale Quantum (NISQ) era \cite{preskill_NISQ_2018}. 
In the standard VQA setting, we minimize the expectation value of the objective function $O$ designed for the target problem with respect to a variational state (ansatz) $\vert \psi(\vec{\theta})\rangle$ prepared by a parameterized quantum circuit (PQC). 
This hybrid scheme is accomplished with a feedback loop between classical computers and quantum devices: the quantum devices repeatedly prepare the variational state $\vert \psi(\vec{\theta}) \rangle$ for estimation of the expectation value of the objective function $\langle \psi(\vec{\theta})\vert O \vert \psi(\vec{\theta})\rangle$, while the classical computers are utilized to optimize the parameters $\vec{\theta}$ based on classical optimization strategies such as gradient descent.

To guarantee that the variational quantum algorithm (VQA) solution is close enough to the exact solution for the target problem, the high expressibility of quantum ansatz $U$ is required. 
In principle, we can increase the depth of the PQC with more quantum gates and training parameters or apply neural network post-processing modules \cite{VQNHE, VQNHE2} to achieve a higher expressibility. 
For VQAs on real quantum hardware, the trainability is also an important factor. Unfortunately, there is a trade-off between expressibility and trainability \cite{PRXQuantum_expressibility_and_trainability} in VQA, and the performance of VQAs is severely limited by optimization issues such as barren plateaus  \cite{mcclean_barren_2018}, which can be induced by the entanglement in the quantum circuit \cite{PRR_entanglement_and_BP, PRL_entanglement_and_BP,PRXQuantum_entanglement_and_BP} and the noise on NISQ devices \cite{wang_noise-induced_BP}. 
The gradients vanish exponentially with the depth of PQC. Therefore, exponential computational resources are required to accurately estimate the gradients value and a large number of iterations is required for the VQA solution to converge. 
Even the gradient-free optimization approaches are also suppressed by the barren plateaus \cite{Arrasmith_gradient_free_BP}. 
In order to mitigate the barren plateaus and achieve better performance from VQA, a series of strategies have been proposed, including parameters initialization methods \cite{Grant_identity_block, Verdon_RNN_initialize, liu_transfer_learning, rad_Bayesian_Learning_Initialization, kulshrestha_beta_distribution, sauvage2021flip_ML, sack2022_WBPs, grimsley_adaptVQE, zhang_gaussian_2022}, local objective function \cite{cerezo_local_poly_vanish,uvarov2021barren_local} and special quantum circuit ansatz \cite{PRX_architecture,Baidu_SEA, bilkis2021semi_VAns, du_quantum_2022_QAS, zhang_jingdong, anand2022information}.

Apart from the barren plateaus, the non-convexity of energy landscapes and the existence of a large number of local minima also strongly limit the trainability of VQAs \cite{VQA_NP} and render training VQAs unscalable. 
The VQA solutions provided by gradient descent can be easily trapped in local minima which are correlated with the initialization position and far away from the global minimum \cite{anschuetz_critical_local_minimum}. 
To escape from the local minima, exponential trials with random initialization of parameters, i.e. exponential optimization trajectories, are required in the general case \cite{Lukin_QAOA}. 
To avoid the local minima, strategies utilize classical neural network \cite{rivera-dean_avoiding_local_minima}, unitary block optimization \cite{PRR_local_minima} and quantum dropout \cite{wang_quantum_2022} have been investigated. To better improve the trainability of VQAs, more effective and universal strategies are still required.

In this Letter, we propose a training algorithm for VQAs with trainable gates activated randomly and progressively to overcome the above challenges. 
We use random activation (RA) to represent this strategy. 
In our new approach, the number of trainable quantum gates is effectively much lower, and thus the barren plateaus can be efficiently mitigated with the expressibility unchanged. 
Additionally, the randomness of activating quantum gates increases the probability of escaping from the local minima. Moreover, the quantum computational resources required as well as the negative effect of quantum noise can be both greatly reduced when experimentally realizing VQA on real quantum hardware.

Without loss of generality, we focus on VQE task in this work, while our approach is applicable to other types of VQAs. 
The VQE has been exploited in a variety of contexts from quantum chemistry \cite{PhysRevX.6.031007_chemistry, kandala_hardware-efficient_2017_chemistry, PhysRevX.8.011021_chemistry, cao2019quantum_chemistry, grimsley_adaptive_2019_chemistry, RevModPhys.92.015003_chemistry, PhysRevA.102.062425_chemistry}, many-body physics \cite{liu_variational_2019, cai_resource_2020, McClean_2016, wecker_progress_2015, 10.21468/SciPostPhys.6.3.029, dallaire-demers_low-depth_2018, tamiya_calculating_2021, uvarov_variational_2020}, to lattice gauge theories \cite{peruzzo_gauge_theory_2014, PRA_gauge_theory} and is specified by a triplet ($O$, $\vert \psi_{i}\rangle$, $U(\vec{\theta})$) including an objective function $O$, an initial quantum state $\vert \psi_{i} \rangle$, and a PQC ansatz $U(\vec{\theta})$. 
The objective function is usually chosen as the expectation or the variance  \cite{zhang_variational_2020, PhysRevB.104.075159, liu_probing_2021} of the system Hamiltonian operator $H$, whose solution gives the ground state or the excited state, respectively. 
There are many choices for PQC structures such as hardware-efficient ansatz (HEA) \cite{HEA},
Hamiltonian variational ansatz (HVA) \cite{PRXQuantum.1.020319_HVA, li_benchmarking_2021_HVA} and the unitary coupled cluster (UCC) ansatz \cite{kutzelnigg_quantum_1982_UCC, kutzelnigg_quantum_1983_UCC, kutzelnigg_quantum_1985_UCC, bartlett_alternative_1989_UCC, taube2006new_UCC, yanai2009accelerating_UCC, yanai2006canonical_UCC, harsha2018difference_UCC}.

{\bf Hamiltonian and circuit ansatz:}
In this Letter, we utilize the new proposed training algorithm to solve the ground state energy problem of the one-dimensional (1D) antiferromagnetic XXZ model, a representative lattice spin model in quantum many-body physics. 
With convincing numerical results and detailed ablation studies, we demonstrate the improved optimization results using our approach and pin down the contributing factors as the mitigation of the barren plateaus and a better chance to reach the global minimum.

\begin{figure*}[t]\centering
	\includegraphics[width=0.9\textwidth]{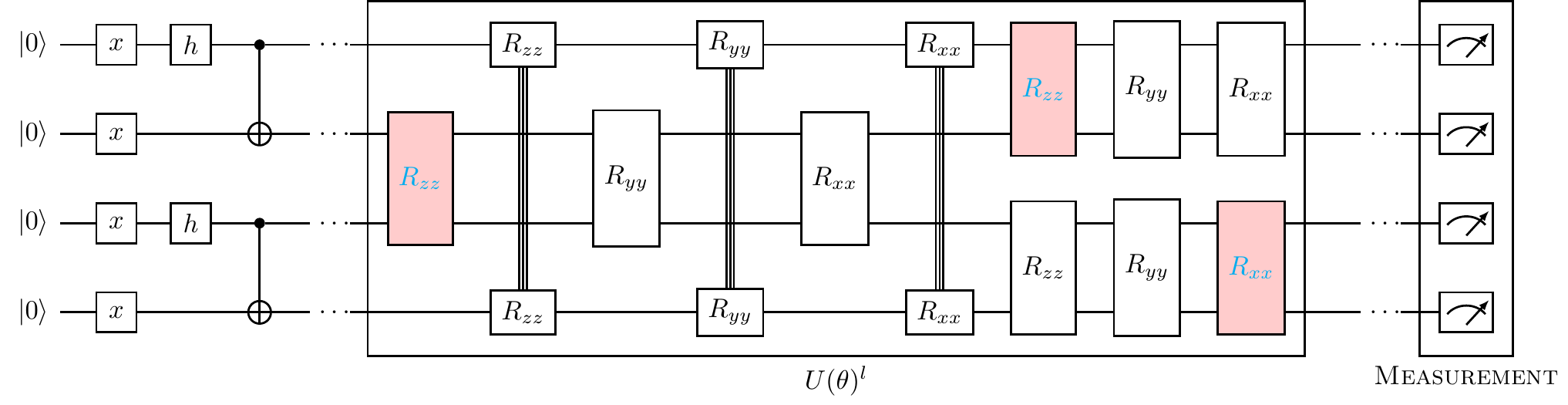}
	\caption{The HVA PQC for XXZ model with $4$ qubits: red rectangles represent the activated two-qubit gates in which the parameters are updated in VQE iterative optimization; white rectangles represent the unactivated two-qubit gates which can be regarded as identity and no need to implement in real experiments. For the training procedure, firstly, we randomly activate $10\%$ two-qubit gates with random initial parameters $\theta_{i} \in [0, 2\pi]$, while other unactivated two-qubit gates' parameters remain as $0$. During training, more and more two-qubit gates are randomly activated and their parameters will unfreeze and update. $l$ is the PQC depth, i.e. how many times the Hamiltonian block repeats.}
	\label{fig:PQC}
\end{figure*}

The Hamiltonian of 1D antiferromagnetic XXZ model with periodic boundary condition reads:
\bea
H = \sum_{i} X_{i}X_{i+1} + Y_{i}Y_{i+1} + J_{z}Z_{i}Z_{i+1},
\eea
where $X,Y,Z$ are Pauli matrices and $J_{z}$ is the $zz$ interaction strength. We choose HVA for the PQC ansatz in this Letter. 
Specifically, a depth-$l$ HVA for XXZ model is shown in Fig. \ref{fig:PQC} and the corresponding circuit unitary is:
\bea
U(\vec{\theta}) = \prod_{i=1}^{l} R^{i, \rm{odd}}_{zz}R^{i, \rm{odd}}_{yy}R^{i, \rm{odd}}_{xx} R^{i, \rm{even}}_{zz}R^{i, \rm{even}}_{yy}R^{i, \rm{even}}_{xx},
\eea
where $R^{i, \rm{odd}}_{\sigma\sigma}  = \prod_{j} \exp(i\theta_{i,2j+1, \sigma}\sigma_{2j+1}\sigma_{2j+2})$ and $R^{i, \rm{even}}_{\sigma\sigma}  = \prod_{j} \exp(i\theta_{i,2j,\sigma}\sigma_{2j}\sigma_{2j+1})$ $(\sigma = x,y,z)$. The parameters $\theta$ of quantum gates are independent tuning parameters.

\begin{figure*}[t]\centering
	\includegraphics[width=0.9\textwidth]{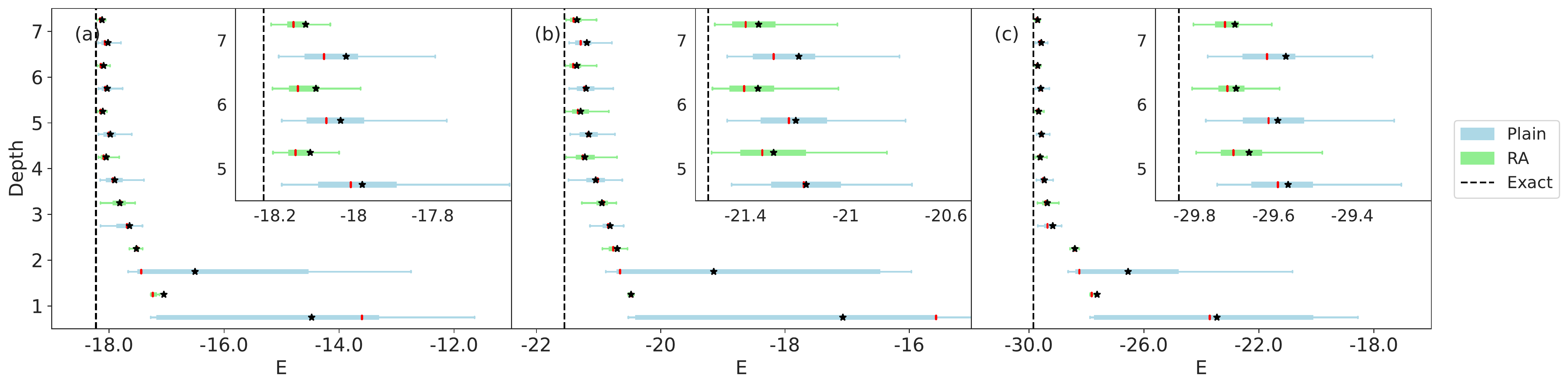}
	\caption{The converged VQE energy in $N=12$ system for the RA training (green) and plain strategy (blue) with different PQC depths $l$: (a), $J_{z}=0.5$; (b), $J_{z}=1.0$; (c), $J_{z}=2.0$. The black star represents the averaged VQE energy and the red line represents the median VQE energy across $500$ independent optimization trials. The inset is the zoom-in of VQE results with deeper PQC. The outliers beyond the caps, which are much larger than the ground truth are not shown (see the Supplemental Material for more details about the standard box plots). The performance of RA training is much better: the mean and median of VQE energy from RA is substantially lower than that obtained from plain training, as more trials are trapped in the local minima for plain training.
	}
	\label{fig:vqe_ds100}
\end{figure*}

Applying the PQC $U(\vec{\theta})$ to the initial prepared state
$\vert \psi_{i} \rangle$ ($\vert \Psi^{-}\rangle$ Bell state, i.e. $\bigotimes_{i=0}^{\frac{N}{2}-1} \frac{1}{\sqrt{2}}(\vert01\rangle-\vert10\rangle)_{2i,2i+1}$), we obtain the output state $\vert \psi(\vec{\theta}) \rangle = U(\vec{\theta}) \vert \psi_{i} \rangle$ and the VQE energy $E(\vec{\theta}) = \langle \psi(\vec{\theta}) \vert H \vert \psi(\vec{\theta}) \rangle$ as the objective function. 
For plain training, all two-qubit gates are activated from the beginning of optimization with random initial parameters $\vec{\theta}$ sampled uniformly from $[0,2\pi]$. 
Then the parameters $\vec{\theta}$ are optimized by gradient descent to reach the lowest energy $\langle \psi(\vec{\theta}^{*}) \vert H \vert \psi(\vec{\theta}^{*}) \rangle$, where $\vec{\theta}^{*}$ are the optimal parameters. 
In principle, a deeper PQC has higher expressibility and a lower converged VQE energy. 
However, barren plateau problems become more severe with large PQC depth, posing a fundamental challenge to identify the optimal parameters $\vec{\theta}^{*}$ as discussed above.

{\bf Training with random gate activation:}
To overcome the challenges presented in the VQA training procedure, we introduce the training algorithm with incrementally random gate activation. Different from the plain training method, only a small fraction of two-qubit gates, for example, $10\%$, are activated for the initial optimization iterations (i.e. the two-qubit gates with the red color shown in Fig. \ref{fig:PQC}). 
Meanwhile, other two-qubit gates stay as identity gates, i.e. $\theta=0$ for these unactivated gates. 
During the optimization, the parameters $\vec{\theta}_{10\%}$ of the initial $10\%$ activated gates will be updated to approach the optimal parameters $\vec{\theta}^{*}_{10\%}$ corresponding to the lowest VQE energy estimation $\langle \psi(\vec{\theta}^{*}_{10\%}) \vert H \vert \psi(\vec{\theta}^{*}_{10\%}) \rangle \le \langle \psi(\vec{\theta}_{10\%}) \vert H \vert\psi(\vec{\theta}_{10\%}) \rangle$. 
Then other random $10\%$ unactivated two-qubit gates are activated. Since the parameters for these newly activated gates are set to $0$ previously, the VQE energy has no sudden change for the incremental activation procedure. 
Now the parameters of the $20\%$ activated gates are optimized together, and this procedure repeats until all parameterized gates are activated, and the optimal parameters $\vec{\theta}_{100\%}^{*}$ are obtained. 
This approach has effectively fewer parameterized gates and can hopefully mitigate the barren plateaus with better VQE performance.

{\bf VQE results comparison:}
We provide numerical analyses on the performance of different training strategies, with numerical implementation based on TensorCircuit package \cite{zhang_tensorcircuit_2022}. 
The results of the plain training and the RA training are shown in Fig. \ref{fig:vqe_ds100} with varying PQC depths and Hamiltonian parameters. 
The results consist of $500$ independent optimization trials on a $12$-qubit system. 
We use Adam \cite{kingma_adam_2017} optimizer with hyperparameters learning\_rate=0.01, decay\_rate=0.9, decay\_steps=100, and the percentage of two-qubit gates activated each time is $10\%$ (results are similar for other reasonable hyperparameter choices and details can be found in the Supplemental Material). We keep the maximal number of iterations maxiter=5000 which is large enough to guarantee that the two different training approaches achieve their best performance. 
This is different from the usual approach where the converge speed instead of the final converged position is the focus \cite{Baidu_SEA, zhang_jingdong, anand2022information}.
We believe our comparison is more suitable to exploit the potential quantum advantage of VQAs.

For the RA training, there are much fewer trials trapped in the bad local minima and the averaged VQE energy is much lower than that obtained from the plain training as shown in Fig. \ref{fig:vqe_ds100}. 
There are two sources for the advantageous performance for the RA training approach, and we will investigate the two factors in detail below: escaping from the local minima more frequently and mitigating the barren plateaus.

The improvement of averaged VQE energy exists for different PQC depths and is particularly significant for shallower quantum circuits with $l=1,2$. 
Note that the barren plateaus are not severe in such a shallow setup. Considering an optimization trajectory of the plain training with random initial parameters $\vec{\theta}$, the converged local minimum of the trajectory is soled determined by the initialization \cite{anschuetz_critical_local_minimum} for plain training. 
And as long as the trajectory is trapped in a local minimum, the VQE performance can be rather bad, providing local minima far away from the exact ground state. 
Instead, if we randomly activate the gate during optimization, for the fixed initial parameters, there are multiple different random optimization trajectories caused by the randomness of how to activate two-qubit gates. 
The extra randomness sources greatly enlarge the parameter space that can be explored for the optimization trajectories, and efficiently help the optimization trajectories escape from local minima more frequently.

\begin{figure}[t]\centering \includegraphics[width=0.4\textwidth]{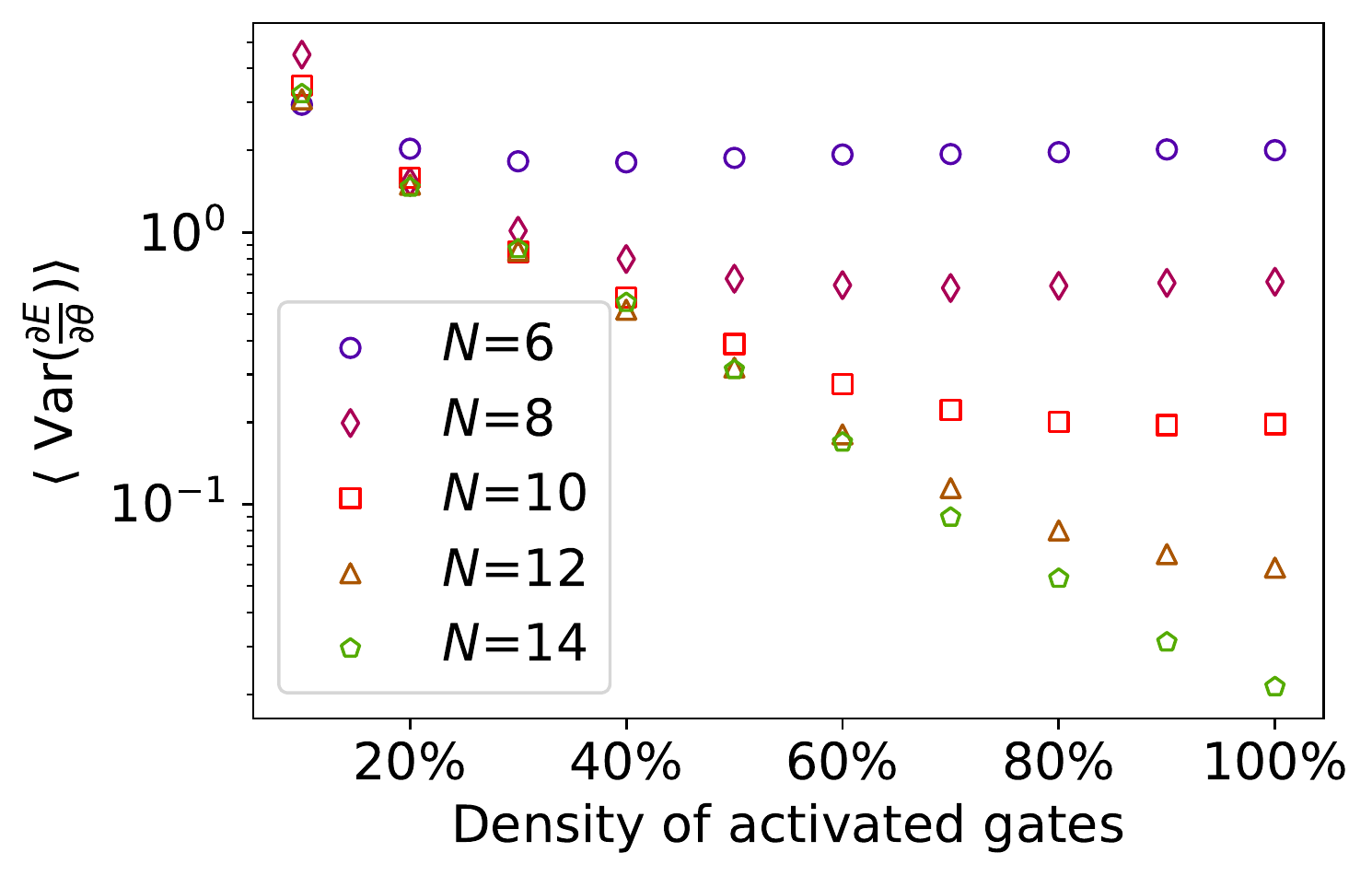}
	\caption{The averaged variance of energy gradients with respect to the parameters of the activated two-qubit gates. The depth of PQC is $l=7$ and the Hamiltonian parameter $J_{z}=1.0$. The barren plateau problems become worse with more activated gates.}
	\label{fig:incremental-var}
\end{figure}

Since the effective circuit depth of our algorithm with only a fraction number of gates is much lower than that of a plain training algorithm, our algorithm also helps to mitigate barren plateaus. 
To observe this quantitatively, the averaged variance of the VQE energy gradients with respect to the parameters of the activated gates are shown in Fig. \ref{fig:incremental-var}. 
With the number of activated gates and the system size increase, the averaged variance decreases exponentially. 
The case of $100\%$ density corresponds to the plain VQE training where all gates are activated. 
And such a plain training strategy suffers from the most severe barren plateau problems. 
As shown in Fig. \ref{fig:incremental-var}, the gradient variance can be efficiently increased during the RA training procedure. 
The mitigation of barren plateaus can also be confirmed via the better median VQE energy of the training. The median energy is not affected by the outlier local minima trials and the improvement on the median is more likely a result from mitigating barren plateaus instead of escaping from a local minimum. 
In sum, the contribution from escaping local minimum can be found via the mean of VQE energy where the distribution of converged energy matters, and the contribution from mitigating barren plateaus can be identified via the median of VQE energy.

{\bf Layerwise activation:} 
We carry out further ablation studies to confirm our analysis on the advantages of RA training. 
A straightforward question is what will happen if there is no randomness in the  gate activation procedure, e.g., activating the two-qubit gates in a layerwise fashion. 
The barren plateaus can also be mitigated due to the effective lower depth for layerwise activation idea. 
But the extra randomness source for escaping from local minima now disappears.
We consider two classes of layerwise optimization: one is the training with layerwise appending activation (LAA), namely, we append the new activated identity layer to the end of the activated layers; the other is the training with layerwise prepending activation (LPA), namely, we prepend the new identity layer to the activated ones.

\begin{figure}[t]\centering \includegraphics[width=0.4\textwidth]{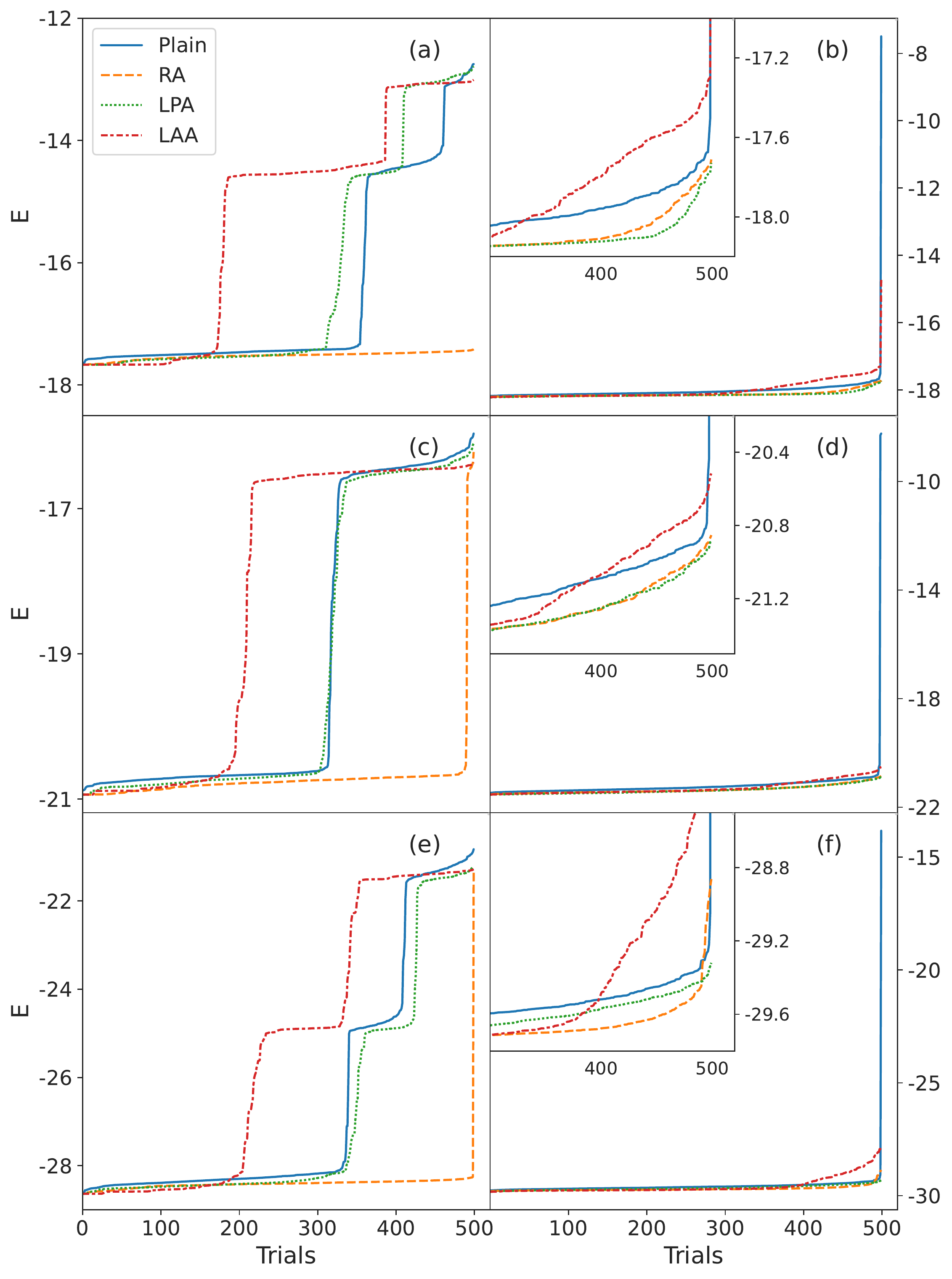}
	\caption{The accumulated distribution of $500$ converged energies from independent optimization trials using four training strategies based on the Adam optimizer with decay\_steps=100 in $N=12$ system: (a), $J_{z}=0.5$, $l=2$; (b), $J_{z}=0.5$, $l=7$; (c), $J_{z}=1.0$, $l=2$; (d), $J_{z}=1.0$, $l=7$; (e), $J_{z}=2.0$, $l=2$; (f), $J_{z}=2.0$, $l=7$. When the quantum circuit is shallow, the VQE performance of RA training is the best because this strategy can escape from local minima more easily. When the quantum circuit is deep, the VQE performance of RA and LPA is very close for $J_{z}=0.5$ and $J_{z}=1.0$ and the VQE performance of RA is the best over $4$ different training strategies for $J_{z}=2.0$.}
	\label{fig:lineplotds100}
\end{figure}

The distribution of the converged VQE results for different training strategies (Plain, RA, LAA, LPA) are shown in Fig. \ref{fig:lineplotds100}. When the quantum circuit is shallow, almost all the trajectories of RA training successfully escape from the local minima, while many trials of the other three training strategies are trapped in local minima and lead to a larger fluctuation in the results due to the absence of the randomness in the activation. 
Therefore, the randomness of activating two-qubit gates is vital and relevant for better escaping from local minima. 
When the quantum circuit is deep, the VQE performance of RA training is still much better than that of the plain training as discussed above. 
We have also observed that the VQE performance of RA and LPA are very close. 
This result indicates that the improvement of VQE performance with deep circuits is mainly due to the mitigation of barren plateaus where RA and LPA are both capable of.

{\bf Entanglement phase transition in RA training:}
A possible reason why RA training works so well even if we consider a fraction of the quantum gates can be demystified from entanglement phase transition perspective. 
We take the HVA circuit structure utilized in this work as the model and the ratio of activated gates $p$ as the driven parameter. 
We can then identify an entanglement phase transition for the output state from an area law to a volume law, similar to the measurement induced entanglement phase transition \cite{QI_MIPT_PRB18, QI_MIPT_PRX19, MIPT_Li_PRB19, MIPT_Clifford_PRL20_Qi, NonlocalMIPT_PhysRevX, jian2021measurement,liuUniversalKPZScaling2022}. 
In our setup, the activation ratio $p$ is relevant, namely $p_c = 0$. 
This fact indicates that even with only 10\% gate activated in this circuit, the output state still lies in the same phase of the full circuit output. 
This entanglement phase transition perspective may partially explain why the RA training is effective. 
See more theoretical and numerical details in the Supplemental Material.

{\bf Discussions:}
We have demonstrated that the RA training can efficiently mitigate the barren plateau problem and help the optimization trajectories escape from local minima based on numerical results and theoretical analysis. 
This new algorithm can be naturally applied to other practical tasks and other flavors of VQAs due to its universal form. 
Additionally, it can be integrated with other strategies \cite{Grant_identity_block, Verdon_RNN_initialize, liu_transfer_learning, rad_Bayesian_Learning_Initialization, kulshrestha_beta_distribution, sauvage2021flip_ML, sack2022_WBPs, grimsley_adaptVQE, zhang_gaussian_2022, cerezo_local_poly_vanish,uvarov2021barren_local, PRX_architecture,Baidu_SEA, bilkis2021semi_VAns, du_quantum_2022_QAS, zhang_jingdong, rivera-dean_avoiding_local_minima, PRR_local_minima, Zhang_2021, Zhang_2022}  to further improve performance and solution quality of VQAs.

From a practical point of view, the new training algorithm brings more benefits when considering implementation on real quantum hardware in the NISQ era. 
In experiments, the circuit gradients are evaluated via parameter shift rule \cite{parameter_shift_1, parameter_shift_2, parameter_shift_3, parameter_shift_4}. 
For a PQC with $p$ training parameters, we have to evaluate the observable expectation on $2p$ sets of parameters. 
In RA training scheme, the effective number of training parameters during the optimization is much smaller than the plain training, which greatly reduces the required number of measurement shots. 
Moreover, the required number of measurement shots for each observable is also lowered due to the mitigated barren plateaus. 
Meanwhile, the number of two-qubit gates required to implement on the hardware in RA training is also lower than the plain training case. 
Therefore, the efficiency in terms of the practical quantum computational resources required is a big advantage for RA training approach. And for the experiments we conducted in this Letter, roughly 500 times less quantum gates are used for RA training (see the Supplemental Material for more detail about the quantum resource efficiency analysis).

In terms of hardware implementation, RA training also has better noise resilience. 
On the one hand, we can stop the activation iteration earlier to achieve a better trade-off between the circuit expressive power and the accumulated quantum noise. 
On the other hand, the effectively shallower circuits also suppress barren plateaus induced by the quantum noise \cite{wang_noise-induced_BP}. 
We show experimental results on noisy VQE settings, and the obtained energies are substantially improved when using RA training (the numerical results and data can be found in the Supplemental Material). 
Therefore, RA training is a practical and standalone training approach that we recommend trying on noisy quantum devices.

{\it Acknowledgements}: We thank Zhou-Quan Wan for helpful discussions. This work is supported in part by the NSFC under Grant No. 11825404 (SXZ, SL, and HY) and by the MOSTC under the MOSTC Grants No. 2021YFA1400100 and No. 2018YFA0305604 (HY). 
S.-K.J is supported by a startup fund at Tulane University.

\bibliographystyle{apsreve}
\bibliography{ref}

\clearpage

\begin{widetext}
	\section*{Supplemental Materials}
	\renewcommand{\thesubsection}{S\arabic{subsection}}
		\setcounter{subsection}{0}
	\renewcommand{\theequation}{S\arabic{equation}}
	\setcounter{equation}{0}
	\renewcommand{\thefigure}{S\arabic{figure}}
	\setcounter{figure}{0}

	\subsection{Box plots}
	In this work, we use box plots extensively to show the distribution of converged energies from different optimization trails. Box plots are used to visualize summary statistics of a dataset, displaying attributes of the distribution like the data's range and distribution.

	In our case, the dataset consists of the converged VQE energies for $N_{t}$ ($N_{t}=500$ for noiseless case and $N_{t}=100$ for noisy case) independent optimization trials. As shown in Fig. \ref{fig:boxplot}, the black star represents the averaged VQE energy and the red line represents the median VQE energy across $N_{t}$ independent optimization trials. The left edge and the right edge of the box correspond to the first quartile ($Q_{1}$) and the third quartile ($Q_{3}$) respectively, where $Q_{1}$ is defined as the median VQE energy between the best VQE energy (minimum) and the median VQE energy,  and $Q_{3}$ is the median VQE energy between the median VQE energy and the worst VQE energy (maximum). $Q_{3}-Q_{1}$ is known as the interquartile range (IQR). After determining the first and third quartiles and the interquartile range as outlined above, then fences are calculated using the following formula:
	\begin{eqnarray}
	\text{Lower fence} = Q_{1} - 1.5 \text{IQR}, \\
	\text{Upper fence} = Q_{3} + 1.5 \text{IQR}.
	\end{eqnarray}
	Any data lying outside these defined bounds can be considered as an outlier. If the best (worst) VQE energy is higher (lower) than the Lower (Upper) fence, the left (right) bar is choosen as the best (worst) VQE energy; otherwise, the left (right) bar is chosen as the best (worst) VQE energy inside the fences. For Plain training, many trials are trapped in the local minima which are higher than the Upper fence and thus not shown in the box plots, namely, the right bar in the box plots corresponds to the worst VQE energy inside the fences instead of the real worst VQE energy. The VQE energy distribution including the worst results trapped in the local minima can be identified in the plots of the accumulated distribution instead, e.g., Fig. \ref{fig:lineplotds100}. On the contrary, due to the large IQR and the ground state energy lower bound, the lower fence always represents the best VQE energy obtained in the experiments.

	\begin{figure}[b]\centering
		\includegraphics[width=0.55\textwidth]{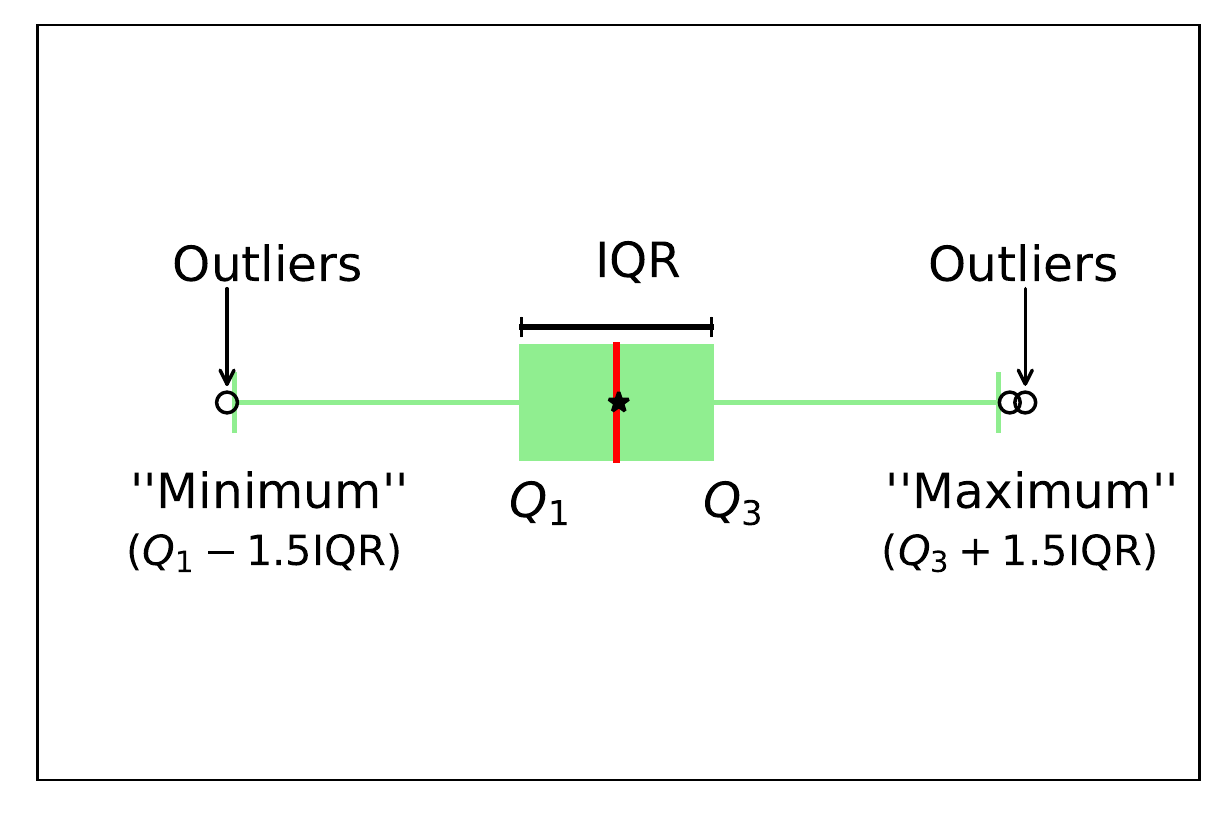}
		\caption{Different parts of a boxplot: the black star represents the averaged VQE energy and the red line represents the median VQE energy across $N_{t}$ independent optimization trials.}
		\label{fig:boxplot}
	\end{figure}

	\subsection{Barren plateaus}
	The averaged gradient variances for different percentages of randomly activated two-qubit gates with $J_{z}=0.5$ and $J_{z}=2.0$ are shown in Fig. \ref{fig:var-0.52.0}. The barren plateau problems become worse with more parameterized quantum gates activated consistent with the results shown in Fig. \ref{fig:incremental-var} with $J_{z}=1.0$.

	\begin{figure}[H]
	\centering
	\begin{minipage}[c]{0.5\textwidth}
	\centering
	\includegraphics[height=5.0cm,width=7.5cm]{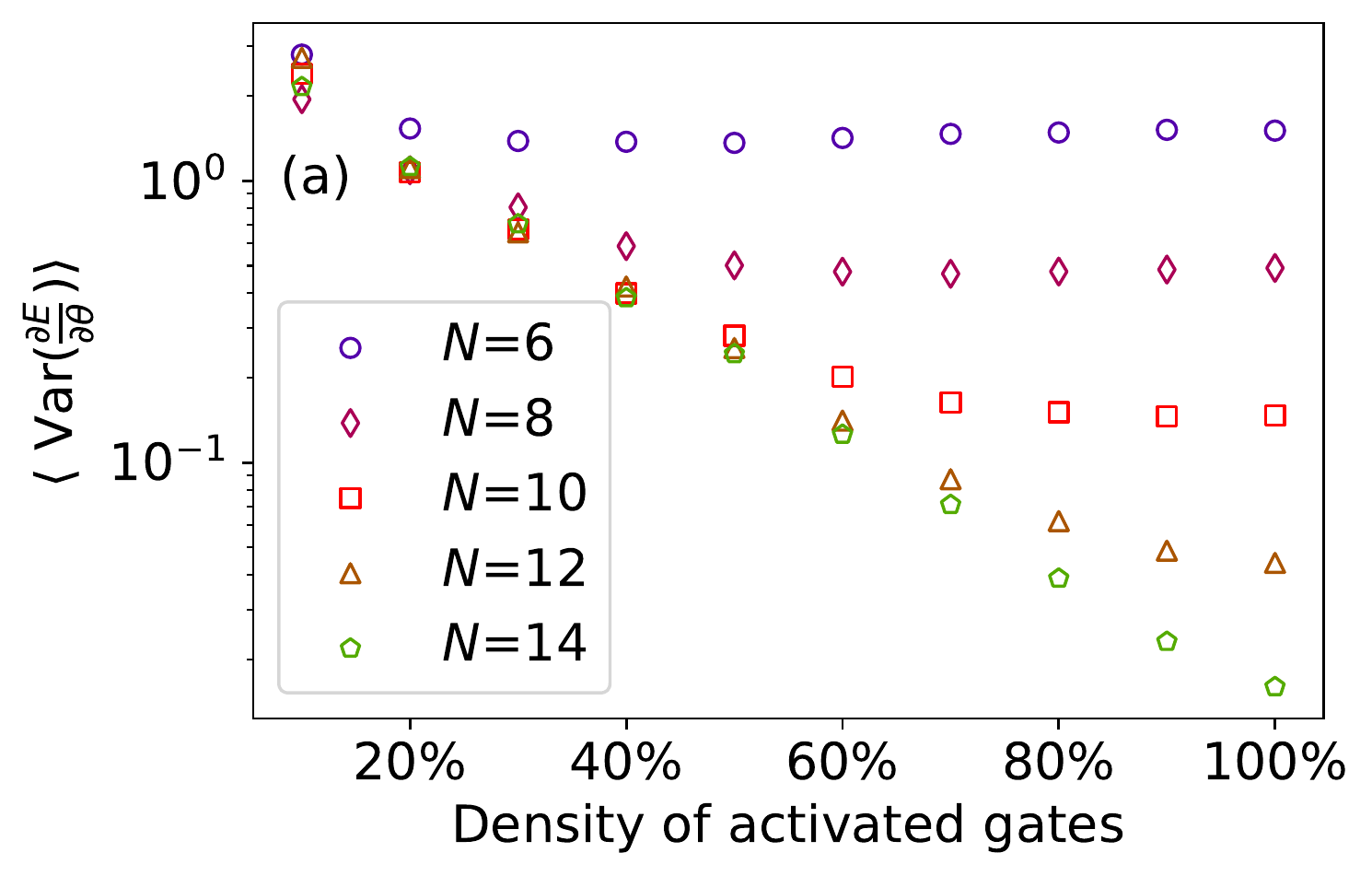}
	\end{minipage}%
	\begin{minipage}[c]{0.5\textwidth}
	\centering
	\includegraphics[height=5.0cm,width=7.5cm]{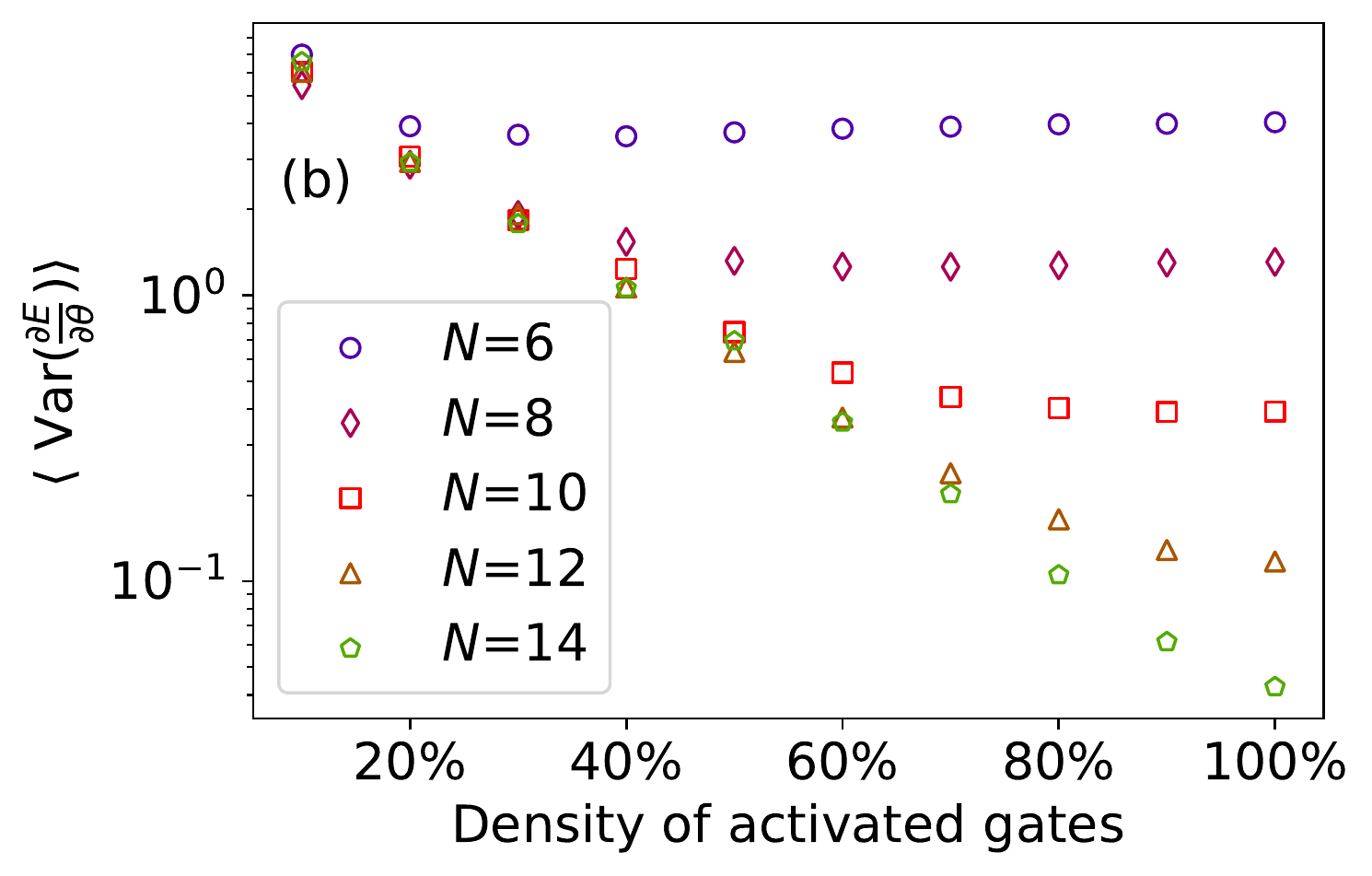}
	\end{minipage}
	\caption{Demonstration of barren plateaus in the VQE problem. The averaged variance of energy gradients with respect to the parameters of the activated two-qubit gates are shown: (a), $J_{z}=0.5$; (b), $J_{z}=2.0$. The PQC depth is $l=7$. The barren plateau problems become worse with more activated gates.}
	\label{fig:var-0.52.0}
	\end{figure}

	\subsection{Best VQE energy analysis}
	
	We also observe that the best VQE energy over $500$ independent trials is similar for two different types of training when the quantum circuit depth is shallow ($l=1,2$), while the best VQE energy of RA training is much lower than that of the plain training when the quantum circuit depth is deep ($l=5,6,7$) as shown in Fig. \ref{fig:vqe_ds100}. Theoretically, the best VQE energy should be the same for different strategies since the PQC ansatz is the same. However, due to the local minima issue, the theoretical global minimum of the ansatz can only be identified with exponential ($O(2^l)$) parameters initialization trials for plain training. When the quantum circuit is shallow, $500$ independent trials are enough to reach this global minimum though many of the trials are trapped in local minima. So the best VQE energy of two training algorithms is close to each other as the theoretical minimum is reached in both cases. However, when the quantum circuit is deep, $500$ independent trials are not enough for plain training to explore the full energy landscape and reach the global minimum. Even if there is a lucky trajectory that does not meet any local minima, the global minimum can not be reached via a large number of iterations due to the vanishing gradients. The improvement of the best VQE energy thus owes to both of the contribution factors of our algorithm.

	\subsection{Adam optimizer and different decap\_steps}
	In the main text, we have fixed the optimizer type and the optimizer's hyperparameters. It is worth noting that the conclusions drawn in the main text don't rely on the special choice of optimizer and its hyperparameters. The results for different decay\_steps and fixed learning\_rate=0.01, decay\_rate=0.9 are shown in Fig. \ref{fig:vqeL7} (the quantum circuit depth is $l=7$). The learning rate on $t$ iterations is given by learning\_rate($t$)=learning\_rate$*$decay\_rate$^{\frac{t}{\text{decay}\_\text{steps}}}$. The choices of decay\_steps influence the decay speed of the learning rate and thus the VQE optimization performance. However, RA training can consistently achieve better performance than that of the plain training for different decay\_steps: the best VQE energy, averaged VQE energy and median VQE energy are all much lower and closer to the ground state energy as shown in Fig. \ref{fig:vqeL7}. The results for different quantum circuit depth $l$ and fixed decay\_steps=80 are shown in Fig. \ref{fig:vqe_ds80}, which also gives similar results as discussed in the main text with decay\_steps=100. With fixed decay\_steps=80, the distribution of the results for different independent trials is shown in Fig. \ref{fig:lineplotds80}.

	\begin{figure}[H]\centering
		\includegraphics[width=0.95\textwidth]{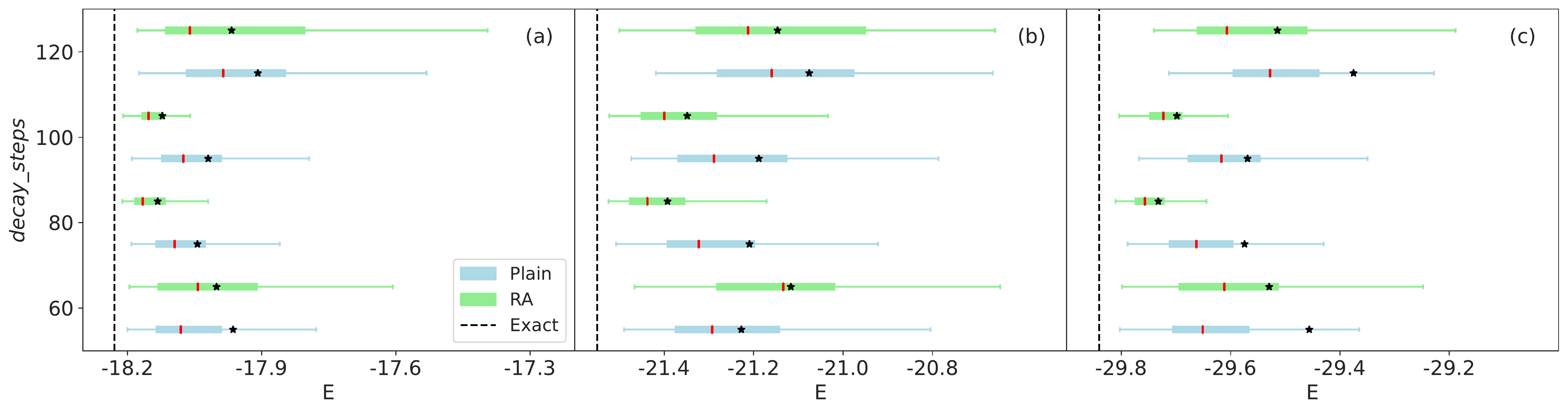}
		\caption{The VQE energy obtained from plain training (blue) and RA training  (green) on a $12$-qubits system with different decay\_steps for Adam optimizer and fixed PQC depth $l=7$: (a), $J_{z}=0.5$; (b), $J_{z}=1.0$; (c), $J_{z}=2.0$. The black star represents the averaged VQE energy and the red line represents the median VQE energy over $500$ independent trials. The performance of RA training is much better than that of the plain training with lower best VQE energy, lower averaged VQE energy, and lower median VQE energy. When $J_{z}=1.0$ and decay\_steps=60, the VQE performance of RA training is worse, which may be caused by the fast-decaying learning rate of the optimizer.}
		\label{fig:vqeL7}
	\end{figure}

	\begin{figure}[H]\centering
		\includegraphics[width=0.95\textwidth]{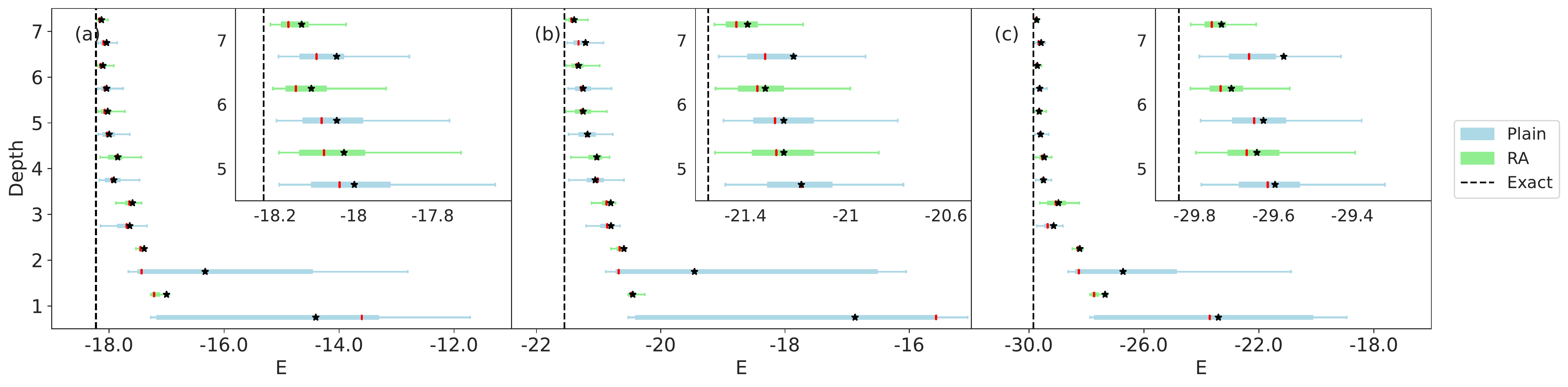}
		\caption{The converged VQE energy for RA training (green) and plain training (blue) with decay\_steps=80 and different PQC depths $l$: (a), $J_{z}=0.5$; (b), $J_{z}=1.0$; (c), $J_{z}=2.0$. The black star represents the averaged VQE energy and the red line represents the median VQE energy across $500$ independent optimization trials. The inset is the zoom-in of VQE results with deeper PQC. The outliers beyond the caps, which are much larger than the ground truth are not shown. 
		The performance of RA training is much better: the averaged VQE energy and median VQE energy from RA training are substantially lower than that obtained from plain training. More trials are trapped in the local minima for plain training.
		The results are similar to those shown in Fig. \ref{fig:vqe_ds100} with decay\_steps=100.}
		\label{fig:vqe_ds80}
	\end{figure}

	\begin{figure}[H]\centering
		\includegraphics[width=0.45\textwidth]{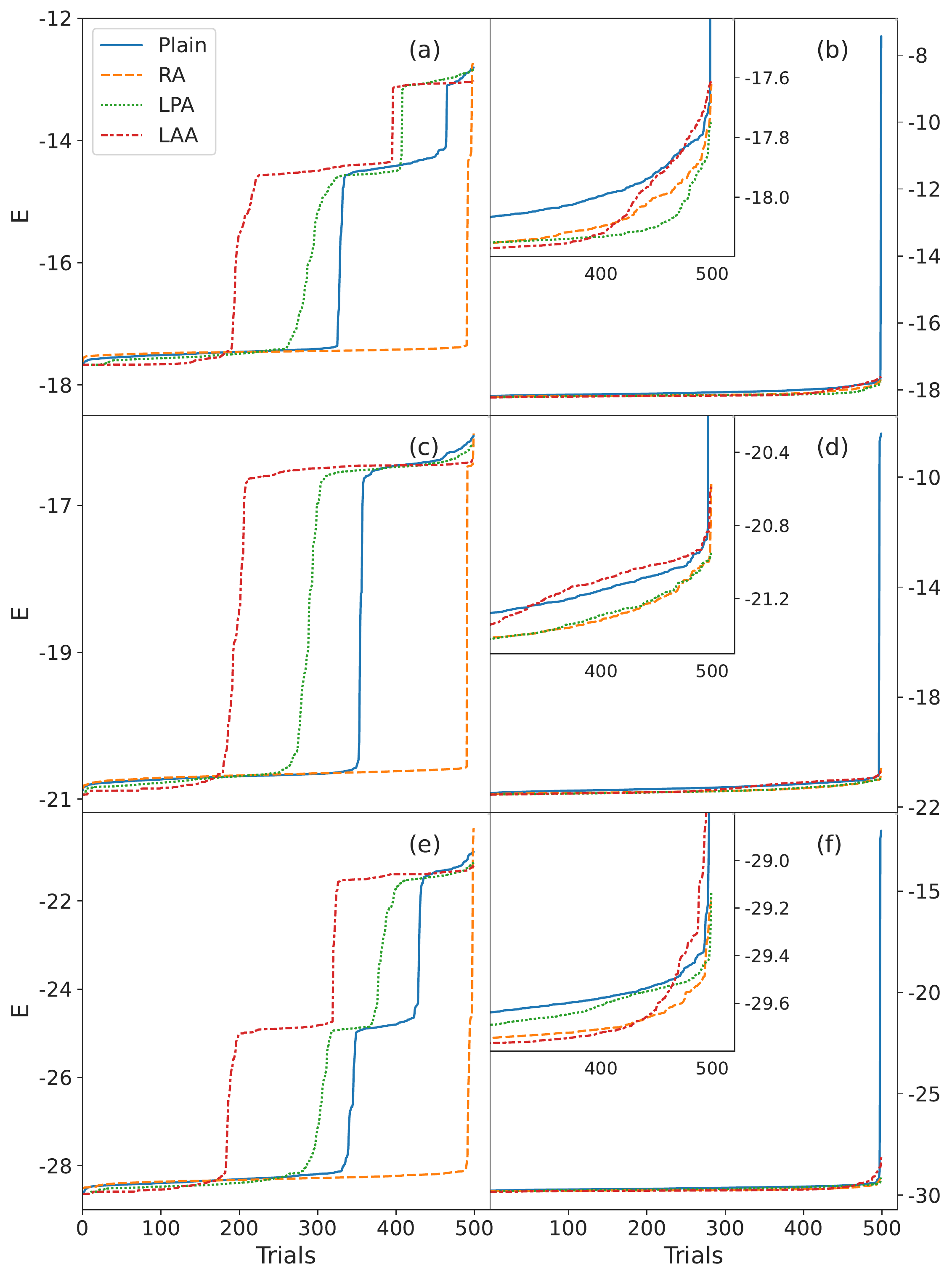}
		\caption{The accumulated distribution of $500$ converged energies from independent optimization trials with four training strategies with decay\_steps=80: (a), $J_{z}=0.5$, $l=2$; (b), $J_{z}=0.5$, $l=7$; (c), $J_{z}=1.0$, $l=2$; (d), $J_{z}=1.0$, $l=7$; (e), $J_{z}=2.0$, $l=2$; (f), $J_{z}=2.0$, $l=7$. When the quantum circuit is shallow, the VQE performance of RA training is the best because this strategy can mitigate barren plateaus and escape from local minima more easily at the same time. When the quantum circuit is deep, the VQE performance of RA and LPA is very close.}
		\label{fig:lineplotds80}
	\end{figure}

	\subsection{SGD optimizer}
	We have also tried different types of gradient descent optimizers such as stochastic gradient descent (SGD) \cite{bottou_stochastic_2012} optimizer to verify the availability of RA training. The results for different $J_{z}$ and different quantum circuit depths $l$ are shown in Fig. \ref{fig:vqe_SGD}. The distribution of the results for different independent trials is shown in Fig. \ref{fig:lineplotSGD}. The conclusions remains similar to the case with Adam optimizer discussed before.

	\begin{figure}[H]\centering
		\includegraphics[width=0.95\textwidth]{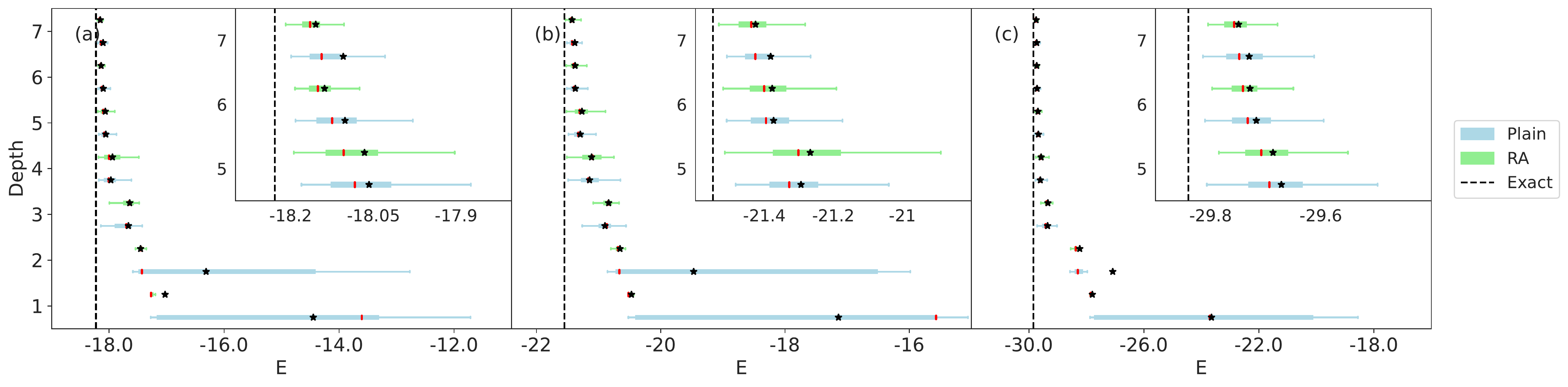}
		\caption{The VQE energy for the plain training (blue) and the RA training (green) with different quantum circuit depths $l$ and fixed learning\_rate=0.01 based on SGD optimizer: (a), $J_{z}=0.5$; (b), $J_{z}=1.0$; (c), $J_{z}=2.0$. The black star represents the averaged VQE energy and the red line represents the median VQE energy. The inset is the zoom-in of VQE results with deeper PQC. The performance of RA training is much better, similar to the case shown in Fig. \ref{fig:vqe_ds100}.}
		\label{fig:vqe_SGD}
	\end{figure}

	\begin{figure}[H]\centering
		\includegraphics[width=0.45\textwidth]{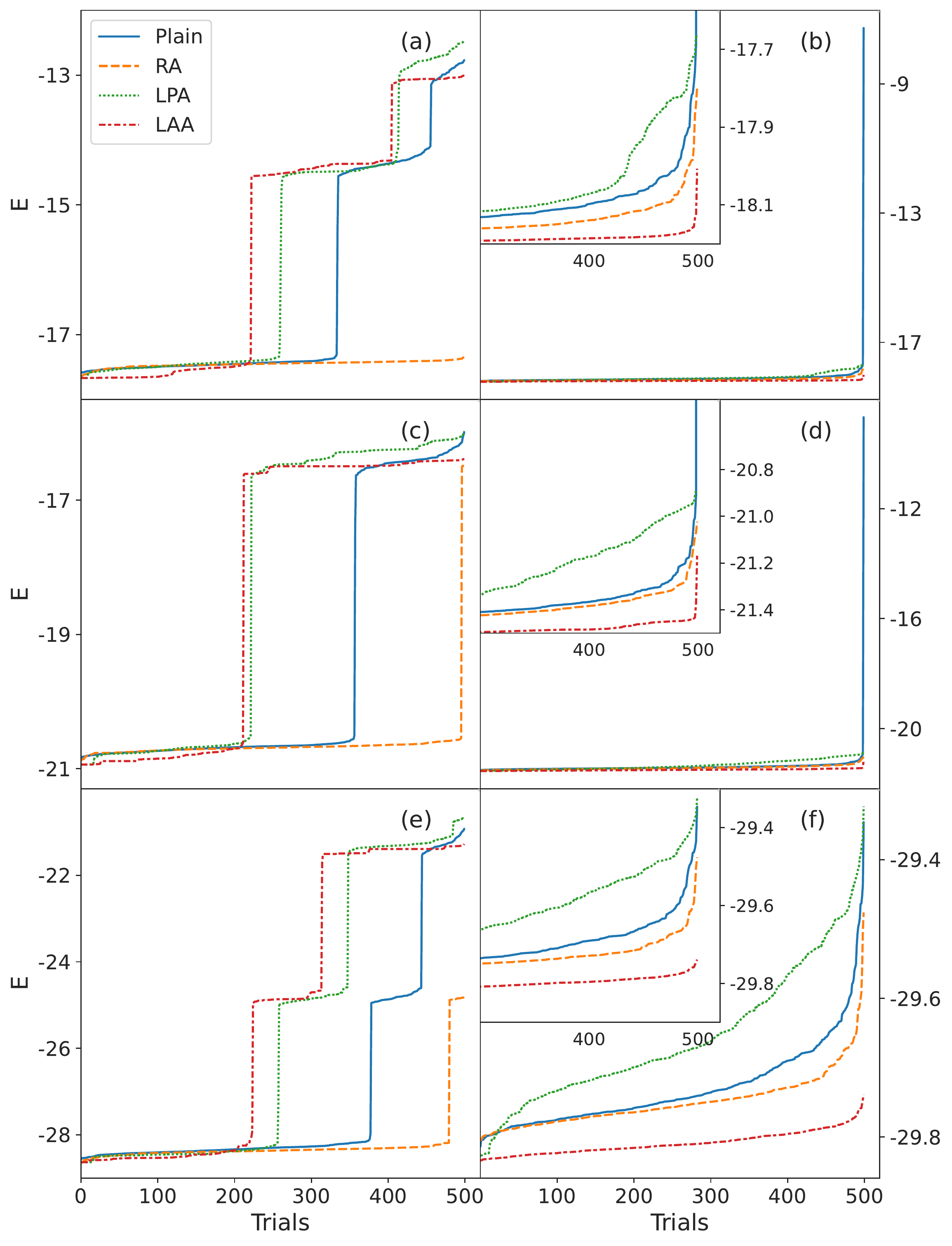}
		\caption{The accumulated distribution of $500$ converged energies from independent optimization trials with four training strategies with learning\_rate=0.01 based on SGD optimizer: (a), $J_{z}=0.5$, $l=2$; (b), $J_{z}=0.5$, $l=7$; (c), $J_{z}=1.0$, $l=2$; (d), $J_{z}=1.0$, $l=7$; (e), $J_{z}=2.0$, $l=2$; (f), $J_{z}=2.0$, $l=7$. When the quantum circuit is shallow, the VQE performance of RA training is the best because this strategy can mitigate barren plateaus and escape from local minima more easily at the same time; when the quantum circuit is deep, it should be noted that the VQE performance of RA is slightly worse than the performance of the LAA, but still better than that of Plain and LPA.}
		\label{fig:lineplotSGD}
	\end{figure}

	\subsection{Different rates of random gate activation}
	The rate to randomly add two-qubit gates each time is $10\%$ in the main text. In this section, we have also tried different rates for RA training. The rate $1$ corresponds to the plain training while the rate $1/m$ indicates that we activate $1/m$ gates each time. The results with fixed PQC depth $l=2$ are shown in Fig. \ref{fig:L2} and the results with fixed PQC depth $l=7$ are shown in Fig. \ref{fig:L7}.

	When the quantum circuit is shallow ($l=2$) and the barren plateau problems are not severe, with the decrease in the rate of adding two-qubit gates, the averaged VQE energy becomes lower and fewer trials are trapped in the local minima. The RA training can help the optimization trajectories escape from the local minima. The best VQE energy and median VQE energy is close with different activation rates because the barren plateau problems are not severe. When this rate is too large, for example, $1/2$, there is no obvious VQE performance improvement between the plain training and the RA training.

	When the quantum circuit is deep ($l=7$) and the barren plateau problems become severe, with the decrease in the rate of adding two-qubit gates, the averaged VQE energy also becomes lower though not as substantial as the shallow circuit case. The fluctuation from different optimization trials is still smaller than that of the plain training. However, the better VQE performance is mainly determined by the mitigation of barren plateaus. The best VQE energy and the median VQE energy of the RA training is closer to the ground state energy with the decrease rate of adding two-qubit gates.

	\begin{figure}[H]\centering
		\includegraphics[width=0.95\textwidth]{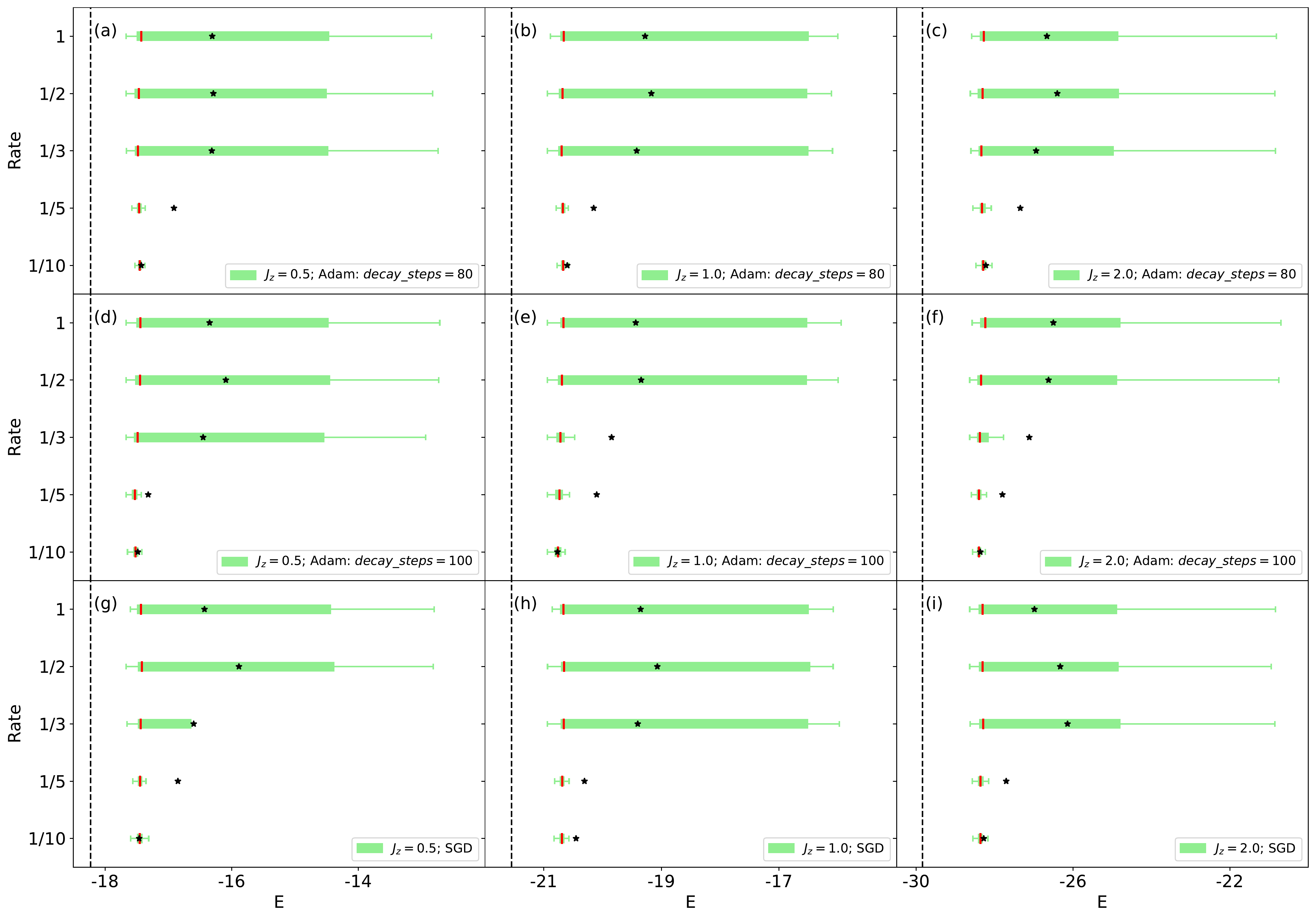}
		\caption{The VQE performance with different rates of adding two-qubit gates each time and fixed PQC depth $l=2$.}
		\label{fig:L2}
	\end{figure}

	\begin{figure}[H]\centering
		\includegraphics[width=0.95\textwidth]{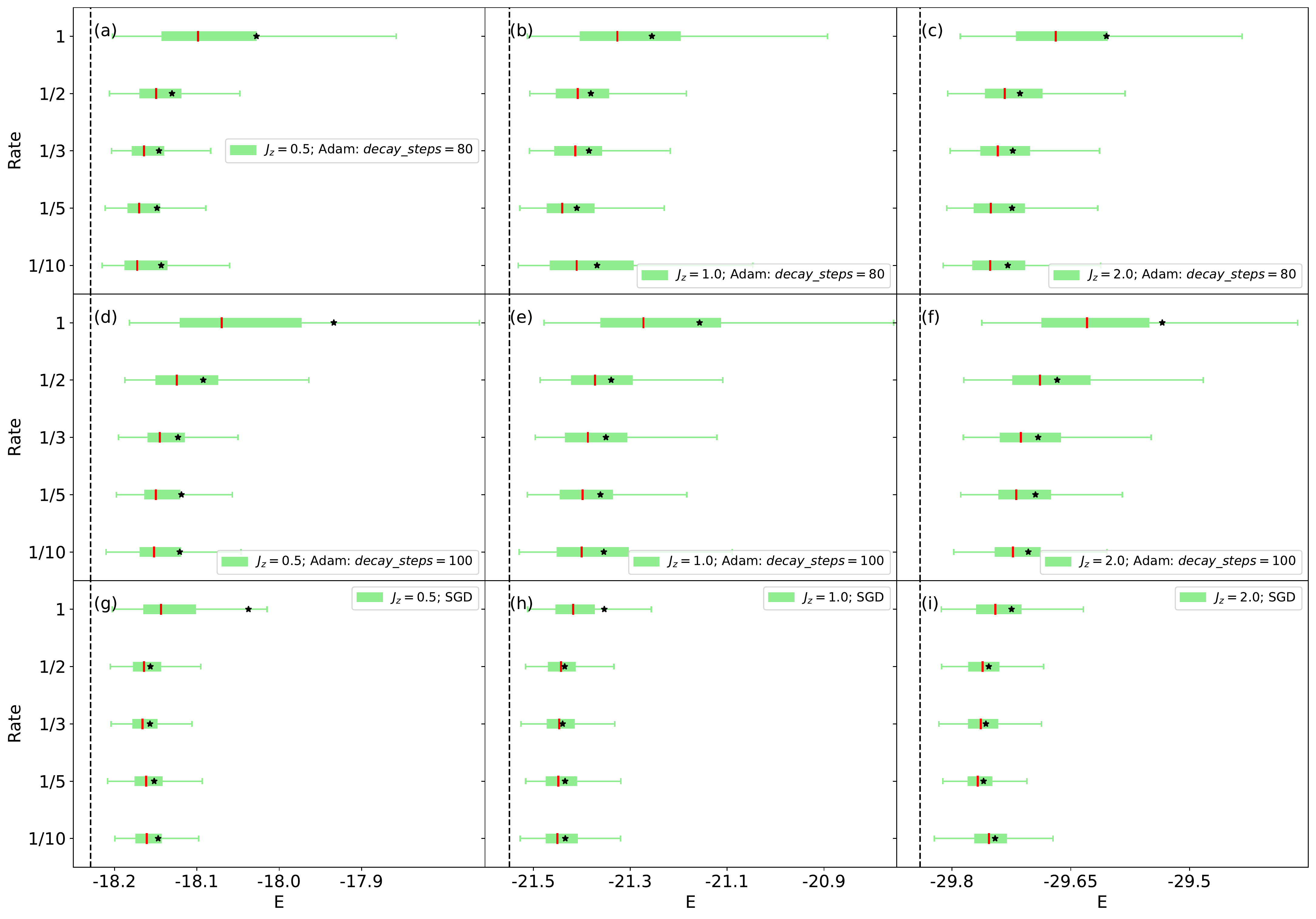}
		\caption{The VQE performance with different rates of adding two-qubit gates each time and fixed PQC depth $l=7$.}
		\label{fig:L7}
	\end{figure}

	In practice, the numerical implementation of RA training regards the gate-adding rate in the sense of statistical average. We use a structure factor to control the activation of each parameterized two-qubit gate. The initial structure factors $\vec{g}$ are sampled uniformly from $(0,1)$. If the structure factor is positive, the two-qubit gate stays as identity gate, i.e. $\theta=0$ for the unactivated gate, and if the structure factor is negative, the two-qubit gate is activated whose parameter $\theta$ will update. Assuming the rate of random gate activation is $1/m$ ($m=10$ in the main text), the structure factors $\vec{g} \mapsto \vec{g}^{\prime} = \vec{g} - \frac{1}{m} \vec{I}$ after one round of random gate activation. Equivalently, $\frac{1}{m}$ fraction of random two-qubit gates activate on average. It is also worth noting that the {\sf vvag} method in TensorCircuit package \cite{zhang_tensorcircuit_2022} admits independent but vectorized optimization loops for batched VQE with different initial parameters $\vec{\theta}$ and structure factors $\vec{g}$ running simultaneously, which unlocks infinite possibilities to explore.

	\subsection{Comparsion with the initialization strategy}
	As discussed in the Refs. \cite{Grant_identity_block, Verdon_RNN_initialize, liu_transfer_learning, rad_Bayesian_Learning_Initialization, kulshrestha_beta_distribution, sauvage2021flip_ML, sack2022_WBPs, grimsley_adaptVQE, zhang_gaussian_2022}, better parameters initialization methods can help VQA reach better performance. In this section, we investigate the plain training where the initial parameters of $10\%$ two-qubit gates are sampled uniformly from $[0, 2\pi]$ and the remaining initial parameters are set to $0$. We use Plain$^{*}$ to represent this strategy. This ablation study is to demonstrate that the gain of RA approach is indeed from the randomness of adding two-qubit gates instead of solely from the better initialization strategy where $90\%$ parameters are set to zero.
	
	The results are shown in Fig. \ref{fig:plainnew}. When the circuit is shallow, more trials of Plain$^{*}$ are trapped in the local minima due to the absence of the randomness of how to add two-qubit gates into the circuit. However, the improvement of the VQE performance is not as pronounced as the case of Plain training with all initial parameters sampled uniformly from $[0, 2\pi]$. Namely, the gain that RA training obtains over plain training consists of both contributions from better initialization strategy and the randomness of gate activation.

	When considering real hardware implementation, compared with Plain$^{*}$, RA can reach a slightly better performance with effectively fewer parameterized gates required, i.e. fewer gradients to be estimated as we discussed in the main text.

	\begin{figure}[H]\centering
		\includegraphics[width=0.95\textwidth]{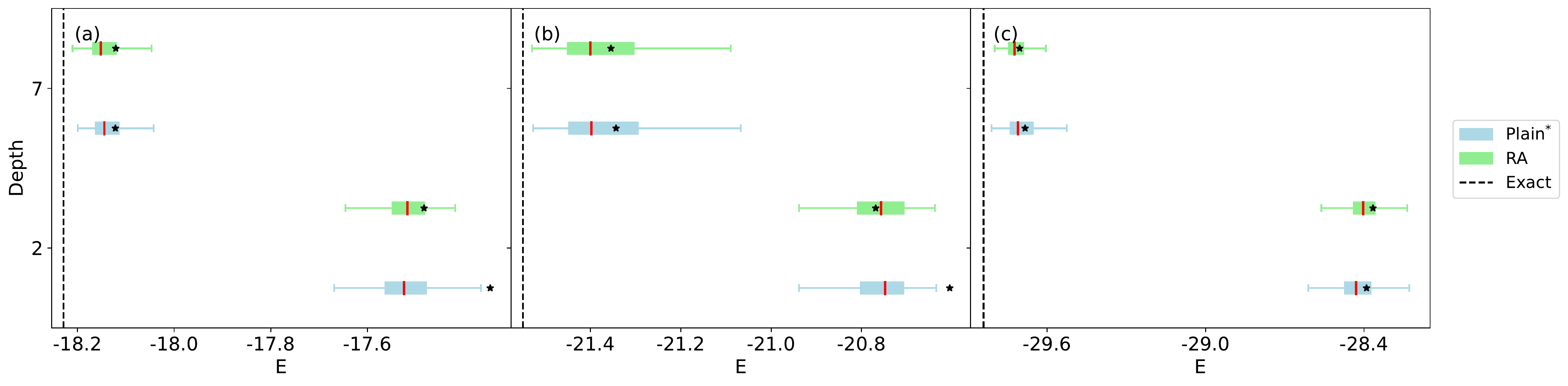}
		\caption{The converged VQE energy ($N=12$, Adam optimizer, decay\_steps=100) for RA training (green) and plain training with better initialization strategy (blue) with different PQC depths $l$: (a), $J_{z} = 0.5$; (b), $J_{z} = 1.0$; (c), $J_{z} = 2.0$. The black star represents the averaged VQE energy and the red line represents the median VQE energy across $500$ independent optimization trials. The results from RA training are slightly better than Plain$^{*}$.}
		\label{fig:plainnew}
	\end{figure}

	\subsection{Robustness against the quantum noise}
	To verify the robustness of this new training algorithm against the quantum noise and the barren plateaus induced by the quantum noise in real hardware, we carry out VQE optimization in the presence of a quantum depolarizing noise with noise strength $p=10^{-3}$ after each two-qubit gate. The depolarizing channel is commonly used in numerical simulation to approximate the noise in a NISQ device. In these experiments, we also use a total iteration step $1000$ instead of $5000$ above, demonstrating that the approach can be much more iteration efficient. The results are shown in Fig. \ref{fig:N6noise}. The quantum noise makes the VQE results much worse. However, RA training can still achieve much better performance than the plain training, in particular with deep PQC in which the plain training totally fails since the noise accumulates with the depth. RA training together with early stop technique indeed provides a feasible and reasonable way to train VQA on noisy devices.

	\begin{figure}[H]\centering
		\includegraphics[width=0.95\textwidth]{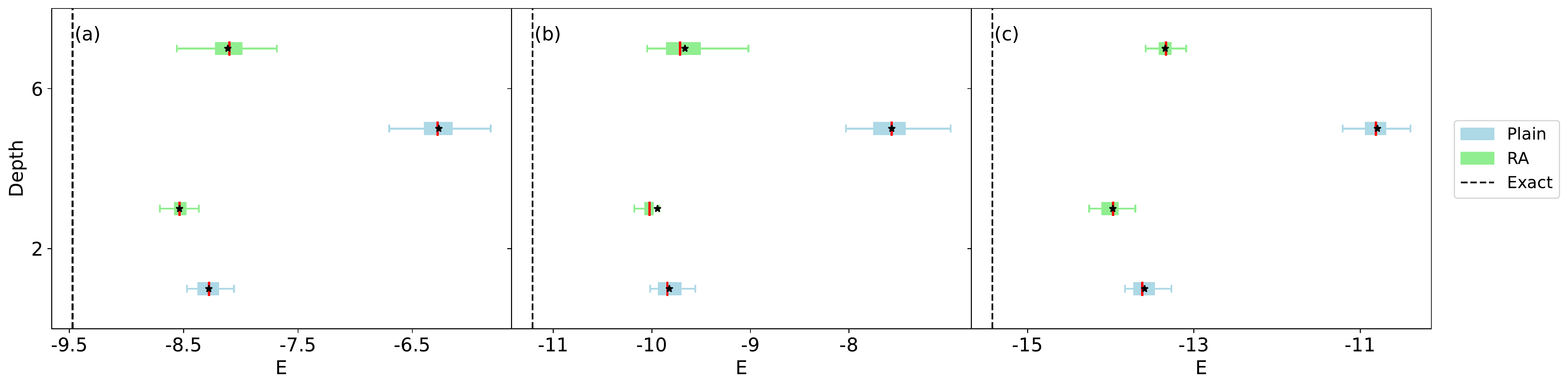}
		\caption{The converged VQE energy ($N=6$, SGD optimizer, learning$\_$rate=0.01) for RA training (green) and plain training (blue) with different PQC depths $l$: (a), $J_{z} = 0.5$; (b), $J_{z} = 1.0$; (c), $J_{z} = 2.0$. The black star represents the averaged VQE energy and the red line represents the median VQE energy across $100$ independent optimization trials. The performance of RA training is much better especially when the circuit depth is large: the averaged VQE energy and median VQE energy from RA training is substantially lower than that obtained from plain training.}
		\label{fig:N6noise}
	\end{figure}

	\subsection{Efficiency comparison for Plain and RA strategies}
	In this section, we will give a rough estimation of the number of two-qubit quantum gates required for Plain and RA strategies in total. And this quantity can be well served as a measure for the total quantum computational resources consumed.

	Assuming $p$ parameters to be optimized in the ansatz (here $p=3Nl$ in our VQE setup), we have to implement the ansatz with $p$ two qubit gates (we omit the constant factor brought by the decomposition of $e^{i\theta\sigma_i\sigma_{i+1}}$ gates). To calculate the circuit gradients via the parameter shift rule \cite{parameter_shift_1, parameter_shift_2, parameter_shift_3, parameter_shift_4}, it requires to evaluate the expectation twice for each trainable parameters. And the accuracy of the gradient for the parameter $\theta$ obtained from $n$ measurement shots is proportional to $\epsilon\sim\frac{1}{\sqrt{N_m}}$ where $N_m$ is the number of measurement shots. Due to the exponentially ($O(e^{-p})$) vanishing variance induced by barren plateaus, we need $O(e^{2p})$ measurement shots to achieve an acceptable accuracy. Recall each measurement shot requires to run a circuit with $p$ quantum gates. Therefore, the number of total quantum gates required in Plain training is:
	\begin{eqnarray}
	N_{g}=2p^{2}e^{2p},
	\end{eqnarray}
	where the factor $2$ corresponds to the two expectations and we have neglected the unimportant constants.
	
	 For RA training, assuming the rate of adding two-qubit gates is $1/m$ ($m=10$ in the main text), the number of required quantum gates is on average:
	\begin{eqnarray}
	N^{\prime}_{g} = \frac{\sum_{k=1}^{m} (\frac{k}{m})^{2}2p^{2}e^{\frac{k}{m}2p}}{m}.
	\end{eqnarray}
	In the $m \longrightarrow \infty$ limit, i.e., we only randomly activate one two-qubit gate each time during the optimization,
	\begin{eqnarray}
	N^{\prime}_{g} &=&  \int_{0}^{1}2p^2 x^2e^{2xp} dx \\ \nonumber
	&=& \frac{-1+(2p^{2}-2p+1)e^{2p}}{2p}.
	\end{eqnarray}
	Then,
	\begin{eqnarray}
	\frac{N^{\prime}_{g}}{N_{g}} = \frac{1}{2p},
	\end{eqnarray}
	in the leading order. Therefore, RA training can have a $2p$ times computational resources efficiency improvement compared with the Plain training (including the Plain$^{*}$ training). In the practical experiments we consider above, $p = 252 (N=12, l=7)$, there is a roughly $500$ times computational resources efficiency improvement. And the computational resources save is more evident when the system gets larger.

	\subsection{Entanglement Phase Transition}
	We focus on the HVA circuit in this work. The circuit consists of blocks of two-qubit gates. We consider the case when only $0\le p\le 1$ ratio of the two qubit gates are activated (selected randomly) and other gates are just identities. These activated two-qubit gates can be regarded as drawn from Haar ensemble or random Clifford ensemble. By measuring the half-chain entanglement entropy of the final output state, we conclude that there is an entanglement phase transition on the random circuit setup driven by the filling ratio $p$.
	
	Fig.~\ref{fig:datacollapse} shows the numerical results on the entanglement phase transition which is numerically simulated using stabilizer circuit, namely, each activated two-qubit gate in the original HVA circuit is replaced with one random two-qubit Clifford gate. And the entanglement of the output state is further averaged over different choice of activated gate and different Clifford implementation of each activated gate. We use $8L$ blocks for the HVA circuit such that the entanglement of the final state is saturated to the infinite time limit. From the results, we clearly see a entanglement phase transition induced by the filling ratio of two-qubit gates $p$, which turns out to be a relevant perturbation. Therefore, we believe that the final state during the RA training is always in the same phase of the final state of plain training. This fact contributes to the effectiveness and correctness of the RA training algorithm.
	\begin{figure}[H]\centering
		\includegraphics[width=0.6\textwidth]{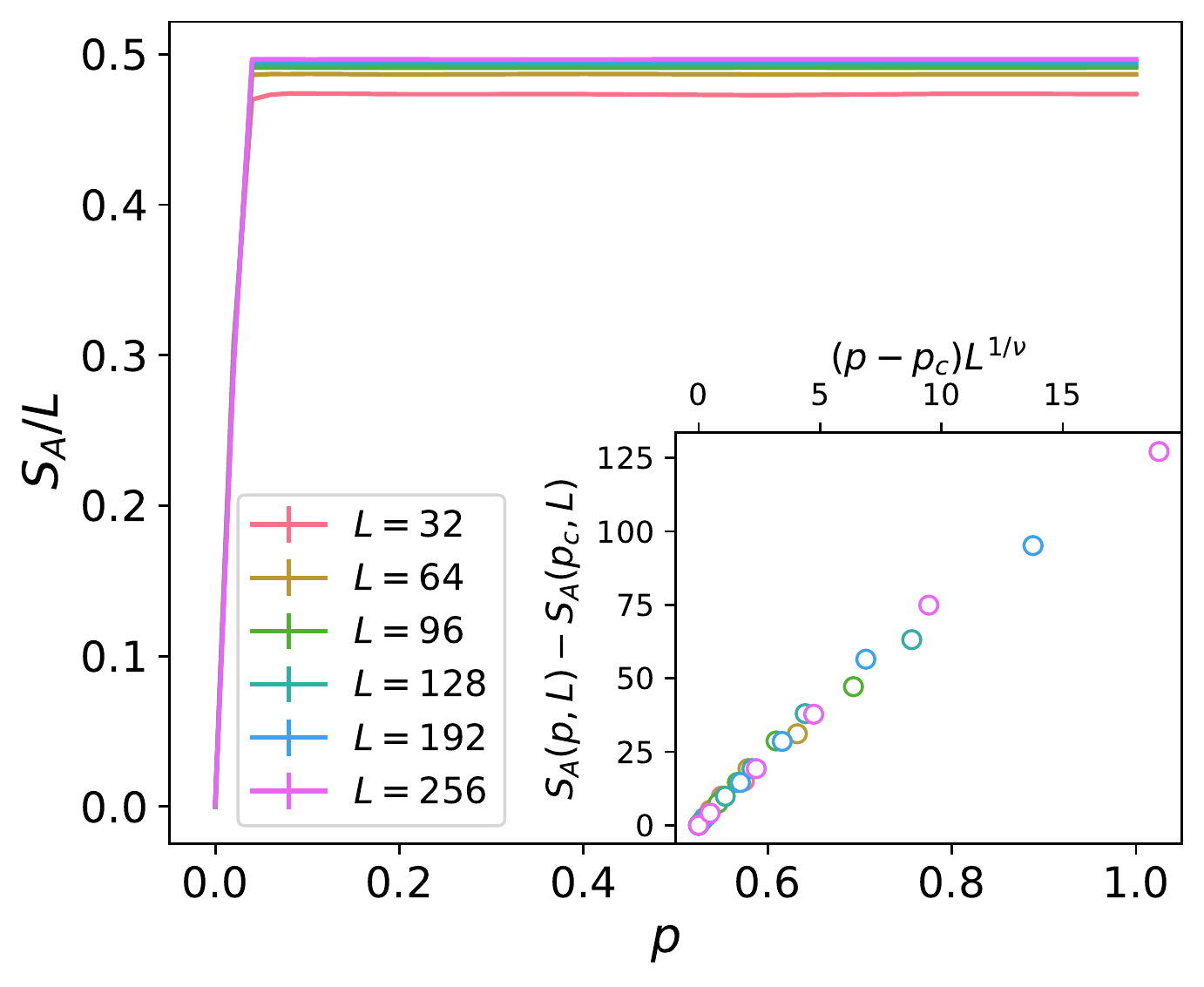}
		\caption{When $p>p_{c}$, the entanglement entropy obeys ``volume law''. The inset is the finite size data collapse with the scaling form: $S_{A}(p)-S_{A}(p_{c}) = F((p-p_{c})L^{1/\nu})$, where $S_{A}(p_{c})=0$. $p_{c}=0$, and the critical exponent is fit to $\nu=0.9$.}
		\label{fig:datacollapse}
	\end{figure}
\end{widetext}
\end{document}